\documentclass[%
groupedaddress,
preprint,
 amsmath,amssymb,
 aps,
 pra,
]{revtex4-1}
\usepackage{graphicx}% Include figure files
\usepackage{dcolumn}% Align table columns on decimal point
\usepackage{bm}% bold math
\usepackage{xcolor}
\begin{document}
\title{Spatial Focusing of Surface Polaritons Based on Cross-Phase Modulation}
\author{Chaohua Tan$^{1,}\footnote {tanch@sdnu.edu.cn}$, Na Li$^1$, Datang Xu$^2$, Zhiming Chen$^3$ and Yong Zhou$^1$}
\affiliation{$^1$School of Physics and Electronics, Shandong Normal University, Jinan 250014, China\\
$^2$School of Electronic and Information Engineering, Changshu Institute of Technology, Changshu 215500, China\\
$^3$School of Science, East China University of Technology, Nanchang  330013, Jiangxi, China}

\date{\today}% It is always \today, today,

\begin{abstract}
We theoretically study the spatial focusing of surface polaritons (SPs) in a negative index metamaterial (NIMM)-atomic gas interface waveguide system, based on cross phase modulation (XPM) in a tripod type double electromagnetically induced transparency (EIT) scheme. In the linear region, we realize the low loss stable propagation of SPs, and the group velocities of the probe and signal fields are well matched via double EIT. In the nonlinear region, we show that giant enhancement of the XPM can be obtained. Using a narrow optical soliton in free space, we realize spatial focusing of the SPs solitons, including bright, multi bright, and dark solitons. The full width at the half-maximum (FWHM) of the SPs soliton can be compressed to about ten nanometers, thus, even nanofucsing can be obtained. The results obtained here have certain theoretical significance for nano-scale sensing, spectral enhancement and precision measurement.
\end{abstract}
\maketitle

\section{INTRODUCTION}{\label{Sec:1}}
Spatial focusing of surface plasmon polaritons (SPPs), especially at nanoscale, recently has been one of the hot spots in the field of micro-nano optics due to its huge application potentials~\cite{Gro 2016}. It not only provides a powerful technical basis for the development of nano optical devices, but also extends the research realm of strong field micro-nano optics~\cite{Dombi 2020}, such as near-field and super-resolution imaging~\cite{Neacsu 2010,Sadiq 2011,Schmidt 2012,Huth 2011, Zhang 2013,Zhong 2017,Liu 2019,Zhu 2019,Lu 2019,Umakoshi 2020,Esmanna 2020}, biological sensing~\cite{Dunn 1999,Anker 2009}, enhanced Raman spectroscopy~\cite{Stockle 2000,Berweger 2010,Bargioni 2011,Stadler 2012,Chen 2014,Lu 2018,Zhang 2018}, nonlinear spectroscopy~\cite{Neacsu 2005,Kauranen 2012,Kravtsov 2016} and photofield electron emission~\cite{Keramati 2020}, {\it etc.}

In 2004, Stockman proposed that SPPs nano-focusing refers to the phenomenon that when SPs propagate along the tapered metallic nanostructure, the propagation energy is highly concentrated at the tip of the tapered structure~\cite{Dombi 2020}. In recent studies, it is mentioned that tapered nanoribbons and metallic tips can be used to construct nano-focusing waveguides by means of micro-nano manufacturing, and have been applied in many fields~\cite{Zia 2006,Verhagen 2007,Choo 2012,Zenin 2015,Li 2019}.
For examples, in 2012, Choo {\it et al.} achieved efficient nano-focusing of SPPs experimentally, which can focus light to a few nanometers with low loss~\cite{Choo 2012}; Zenin {\it et al.} used the tapered nanoribbon structure to detect that the light field energy can be concentrated in a space of tens of nanometers through nanofocusing, and the near field intensity at the tip of the tapered waveguide can be enhanced to the order of thousands in 2015~\cite{Zenin 2015}; Zhu {\it et al.} realized SPPs nano-focusing at the tip of the round tower structure, which enhanced the electric field at the tip of the round tower and obtained nano-level light spots in 2019~\cite{Zhu 2019}. SPPs nano-focusing can also be used as an alternative method to prepare nano-light sources for optical nanoimaging~\cite{Umakoshi 2020,Esmanna 2020}. For example, in 2020, Umakoshi {\it et al.} used nano-focusing of SPPs on tapered metallic nanostructures with a tip diameter of tens of nanometers to obtain a white nano light source in the entire visible light wavelength range~\cite{Umakoshi 2020}.

However, metallic nano structures adopted in the above research are all based on high precision micro/nano manufacturing technology, once the structures is prepared, the performance of the device is almost determined, and it is lack of active control. In this work, we propose an active approach to achieve spatial focusing of surface polaritons (SPs, which is excited at the surface of negative index metamaterials and can propagate with low loss for a long distance~\cite{Kamli 2008,Zorgabad 2018,Zorgabad 2019,Liu 2020}) based on cross-phase modulation (XPM). Actually, the research of compressing pulses in time domain or frequency domain via XPM in fiber optics is very mature~\cite{Agrawal 1990,Agrawal book 2019,Liu 2010,Liu 2011}, the physical mechanism is based on the competitive interaction between dispersion and nonlinearity. Thus, we can also use the competitive interaction between diffraction and nonlinearity to realized the spatial focusing of SPs. Such expansion is like the relation between temporal soliton and spatial soliton~\cite{Kivshar 2003}.

In this article, we propose a general theoretical scheme to investigate spatial focusing of SPs in a negative index metamaterial (NIMM)-atomic gas interface waveguide, based on XPM in a tripod type double electromagnetically induced transparency (EIT) system.
First, we obtain the low loss stable propagation of SPs, and the group velocities of the probe and signal fields are well matched under the double EIT condition in the linear region, and then, giant enhancement of the XPM can be obtained in the nonlinear region. Finally, the coupled NLSEs are derived in our system, by adopting the bright-bright soliton pair, multi bright solitons pairs and dark-dark soliton pair solution as the initial condition, we realize the spatial focusing of SPs solitons via XPM between the narrow optical soliton in free space and the SPs soliton, and even nanofocusing.

The rest of the paper is organized as follows: In Sec.~\ref{Sec:2}, we propose the theoretical model for the study. In Sec.~\ref{Sec:3}, the linear and nonlinear properties of the signal and probe fields, together with the nonlinear envelope equantions are given. In Sec.~\ref{Sec:4}, spatial focusing of SPs is studied. Finally, in Sec.~\ref{Sec:5}, we summarize the main work of this paper.

\section{THEORETICAL MODEL}{\label{Sec:2}}
 %%%%%%%%%%%%%%%%%%%%%%%%%%%%%%%%%%%%%%%%%%%%%%%%%%%%%%%%%%%%%%%%%%%
 \begin{figure}
\centering
\includegraphics[scale=0.8]{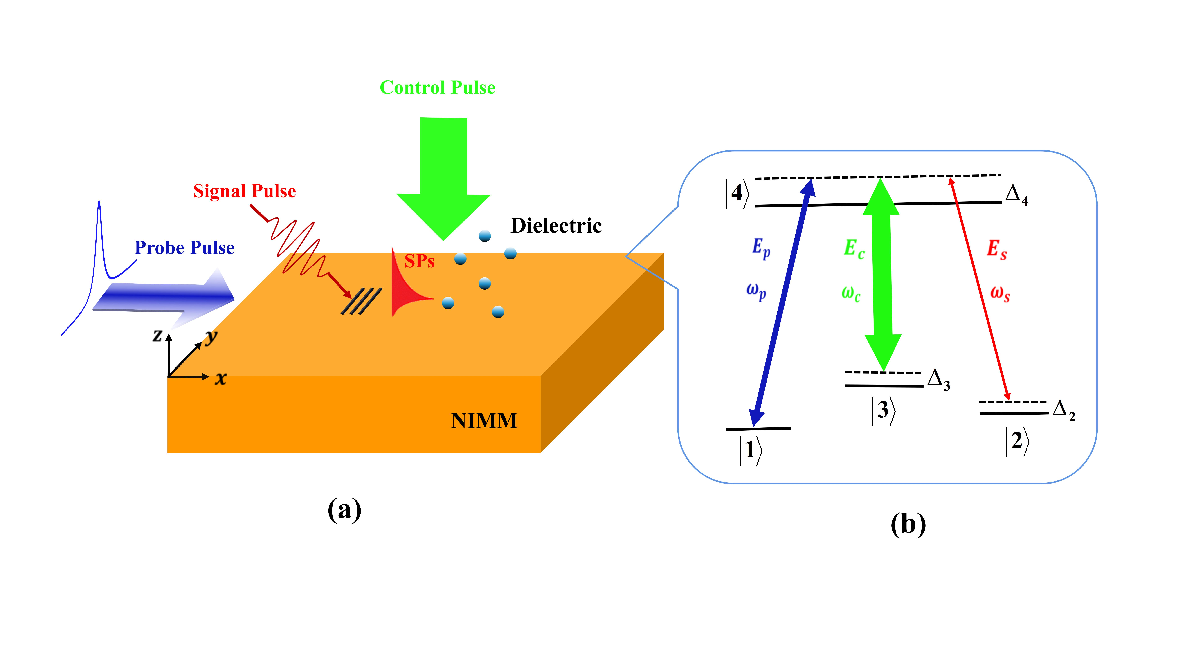}
\caption{\footnotesize (Color online) (a) A scheme for nanofocusing of the SPs via XPM. The SPs is excited and propagates in the $x$ direction. The probe field which is narrow in the $y$ direction propagates along the surface of the NIMM in the $x$ direction. The control field incidences in the vertical direction. (b) The cold atomic gas is charged above the surface, and with a tripod-type four-level excitation configuration. The probe, signal and control fields are coupled to the transition $|1\rangle\leftrightarrow|4\rangle$, $|2\rangle\leftrightarrow|4\rangle$ and $|3\rangle\leftrightarrow|4\rangle$, respectively. $\Delta_j~(j=2, 3, 4)$ are the optical detunings.}
\label{Fig:1}
\end{figure}
%%%%%%%%%%%%%%%%%%%%%%%%%%%%%%%%%%%%%%%%%%%%%%%%%%%%%%%%%%%%%%%%%%%%%%%%%
%
The system under study consists of a layer of NIMM in the lower half plane $z<0$ and a cold atomic gas in the upper half plane $z>0$, as shown in Fig.\ref{Fig:1}(a). The permittivity $\varepsilon_1$ and permeability $\mu_1$ of the NIMM are given by the Drude model in optical region~\cite{Kamli 2008}. The SPs is excited and propagates in the $x$ direction. The probe field which is narrow in the $y$ direction propagates along the surface of the NIMM in the $x$ direction. The control field incidences in the vertical direction.

The cold atomic gas is chosen as a double-EIT excitation medium with a tripod-type four-level configuration. The three fields are interacting with the atoms coherently as shown in Fig.\ref{Fig:1}(b). The weak probe field $\mathbf{E}_p$ with center angular frequency $\omega_p$ couples $|1\rangle\leftrightarrow|4\rangle$, the weaker signal field $\mathbf{E}_s$ with center angular frequency $\omega_s$ couples the transition $|2\rangle\leftrightarrow|4\rangle$, and the strong control field $\mathbf{E}_c$ with center angular frequency $\omega_c$ couples $|3\rangle\leftrightarrow|4\rangle $, $\Delta_j$ ($j=2,3,4$) are the optical detunings. The atoms occupying the excited state $|4\rangle$ can spontaneously radiate to the three ground states ($|1\rangle$, $|2\rangle$, and $|3\rangle$) with spontaneous emission rates $\Gamma_{j4}$ $(j=1,2,3)$. The transition between the three ground states are forbidden, but, there may be other dephasing processes, such as collision, we denote such dephasing rates by $\Gamma_{jl}$ (between the states $|j\rangle$ and $|j\rangle$, $j,l=1,2,3$), those dephasing rates are relatively small comparing with spontaneous emission rates, thus, we assume such processes could not cause the population exchange between the ground states.
Such a system can support the propagation of lossless SPs, and provides a great platform for studying the nonlinearity of SPs during interacting with the coherent medium~\cite{Kamli 2008}.

The signal field is chosen as a TM mode of the waveguide, with the electric field being $\mathbf{E}_s=\mathcal{E}_s\mathbf{u}_s(z)e^{i(k(\omega_s)x-\omega_st)}+c.c.$, in which, $\mathbf{u}_s(z)$ is the mode function in the $z$ direction~\cite{Liu 2020}. The probe and control fields are chosen as $\mathbf{E}_{c(p)}=\mathcal{E}_{c(p)}\mathbf{e}_{c(p)}e^{i(k_{c(p)} x-\omega_{c(p)}t)}+c.c.$. $\mathcal{E}_j$~$(j=p, s, c)$ represents the envelope of the three fields.

In interaction picture, under the electric-dipole and rotating-wave approximations, the Hamiltonian of the system reads
\begin{equation}\label{H}
\hat{H}_{int}=-\hbar\sum_{j=1}^{4}\Delta_j|j\rangle\langle j|-\hbar[\Omega_c|4\rangle\langle3|+\Omega_p|4\rangle\langle1|+\zeta_s(z)\Omega_se^{i\theta_s}|4\rangle\langle2|+h.c.],
\end{equation}

with $\Omega_c=|\mathbf p_{34}|\mathcal{E}_c/\hbar$, $\Omega_p=|\mathbf p_{14}|\mathcal{E}_p/\hbar$, $\Omega_s=|\mathbf p_{24}|\mathcal{E}_s/\hbar$ being the half-Rabi frequencies of the control, probe and signal fields, respectively. $\mathbf p_{jl}=p_{jl}\mathbf{e}_{jl}$ is the electric dipole matrix element related to the transition from $|j\rangle$ to $|l\rangle$. $\zeta_s(z)=\mathbf u_s(z)\cdot\mathbf e_{24}$ is the mode intensity distribution function describing the interaction weight between SPs and atoms along the $z$ firection. $\theta_s=[k(\omega_s)+k_2-k_4]$ are phase mismatches caused by the eigen dispersion of SPs, where $k_l~(l=2,~4)$ refers to the wave number of the state $|l\rangle$.

The interaction information of the system is given by the density matrix $\sigma$, which is a $4\times4$ matrix, and the evolution of $\sigma$ is governed by the optical Bloch equation~\cite{Liu 2020}
\begin{equation}\label{bloch}
\frac{\partial \sigma}{\partial t}=-\frac{i}{\hbar}[\hat{H}_{int}, \sigma]-\Gamma\sigma,
\end{equation}
with $\Gamma$ being a $4\times4$ relaxation matrix which describes the spontaneous emission and other dephasing effects of the system. The detailed expressions of the density matrix $\sigma$ are given in Appendix~\ref{AppendixA}.

Under the condition of slowly varying envelope approximation, the Maxwell equations can be reduced to
\begin{subequations}\label{M}
\begin{eqnarray}
&&i(\frac{\partial}{\partial x}+\frac{1}{c}\frac{\partial}{\partial t})\Omega_p+\frac{1}{2k_p}\frac{\partial^2}{\partial {y^2}}\Omega_p+\kappa_{14}\sigma_{41}=0,\label{Op}\\
&&i(\frac{\partial}{\partial x}+\frac{1}{n_{\rm eff}}\frac{1}{c}\frac{\partial}{\partial t})\zeta_s(z)e^{i\theta_s}\Omega_s+\frac{1}{2k(\omega_s)}\frac{\partial^2}{\partial y^2}\zeta_{s}(z)e^{i\theta_s}\Omega_s+\kappa_{24}\sigma_{42}=0, \label{Os}
\end{eqnarray}
\end{subequations}
where $\kappa_{14}={\cal N}_a\omega_{p}^2|\mathbf{p}_{14}|^2/[{2\hbar\varepsilon_0 c^2\tilde{k}(\omega_p)}]$ and $\kappa_{24}={\cal N}_a\omega_{s}^2|\mathbf{p}_{24}|^2/[{2\hbar\varepsilon_0 c^2\tilde{k}(\omega_s)}]$ are the coupling coefficients of the probe field and the signal field, $\tilde{k}={\rm Re}(k)$, $n_{\rm eff}=k(\omega_s)c/\omega_s$ is the effective refractive index,  ${\cal N}_a$ represents the number density of the atoms, and $c$ is the speed of light in vacuum.

Equations~(\ref{bloch}) and~(\ref{M}) can totally describe the interaction and propagation properties of our system, which is known as the Maxwell-Bloch (MB) equations. Note that, the Rabi frequencies of the three fields in our system satisfies the condition $|\Omega_s|\ll|\Omega_p|\ll|\Omega_c|$, thus, Eqs.~(\ref{bloch}) and (\ref{M}) can be solved by the multi-scale method~\cite{Huang 2005}.

\section{THE LINEAR AND NONLINEAR PROPERTIES OF THE SYSTEM }{\label{Sec:3}}
Firstly, we make the asymptotic expansion: $\sigma_{jl}-\sigma_{jl}^{(0)}=\epsilon\sigma_{jl}^{(1)}+\epsilon^2\sigma_{jl}^{(2)}+...$, $\Omega_p= \epsilon\Omega_p^{(1)}+\epsilon^2\Omega_p^{(2)}+\epsilon^3\Omega_p^{(3)}+...$ and $\Omega_s= \epsilon^2\Omega_s^{(2)}+\epsilon^3\Omega_s^{(3)}+...$, where, $\epsilon$ is a dimensionless small parameter characterizing the typical amplitude ratio of the probe field and the signal field, and $\sigma_{jl}^{(0)}$ is the initial state of the system, and all the physical quantities on the right side of the equation are functions of the multiple scales variable $t_j=\epsilon^jt$, $x_j=\epsilon^jx(j=0,2)$ and $y_1=\epsilon y$. We can get a series of linear but nonhomogeneous equations about $\sigma_{jl}^{(\alpha)}$, $\Omega_p^{(\alpha)}$, and $\Omega_s^{(\alpha)}$ , that can be solved order by order.

\subsection{LINEAR PROPERTIES OF THE SYSTEM}

Firstly, we will show that a ultra-low loss propagation of SPs in the linear region, and group velocity matching between the probe and signal fields can be realized via the double EIT effect.

When there is no probe field and signal field, we can obtain the zero-order solution of the system, which corresponding to the initial state of the system, $\sigma_{jl}^{(0)}=0$, $\sigma_{33}^{(0)}=\sigma_{44}^{(0)}=0$, $\sigma_{11}^{(0)}+\sigma_{22}^{(0)}=1$. Thus, there are no coherence between the states, and all atoms populate on the ground states $|1\rangle$ and $|2\rangle$. For convenience, we have assumed that $\sigma_{11}^{(0)}=\sigma_{22}^{(0)}=1/2$ in the following discussions, which means atoms is uniformly distributed in the states $|1\rangle$ and $|2\rangle$.

When turning on the probe and signal fields, we can obtain the first and second order solutions of the system, respectively, and both are linear problems. We assume that the probe field (signal field) $\Omega_p^{(1)}$ $(\Omega_s^{(2)})$ is proportional to $ e^{i\theta_p}$ $(e^{i\theta_s})$ with $\theta_p=K_p(\omega)x_0-\omega t_0$ $(\theta_s=K_s(\omega)x_0-\omega t_0)$, which means $\Omega_p^{(1)}=F_1e^{i(K_p(\omega)x_0-\omega t_0)}$ $(\Omega_s^{(2)}=F_2e^{i(K_s(\omega)x_0-\omega t_0)})$, where $F_1$ ($F_2$) are slowly varying envelop function of multiscale variables yet to be determined. Then, we can obtain the linear dispersion relations of the probe field and signal field interacting with the double EIT medium, which read
\begin{subequations} \label{K}
\begin{eqnarray}
&&K_p(\omega)=\frac{\omega}{c}+\kappa_{14}\frac{(\omega+d_{31})\sigma_{11}^{(0)}}{D_p}\label{K_p},\\
&&K_s(\omega)=\frac{1}{n_{\rm eff}}\frac{\omega}{c}+\kappa_{24}\frac{(\omega+d_{32})\sigma_{22}^{(0)}}{D_s}\label{K_s},
\end{eqnarray}
\end{subequations}
where $\omega$ is the frequency shift to the center frequency of the probe and signal fields, and we have defined $D_p=|\Omega_c|^2-(\omega+d_{31})(\omega+d_{41})$ and $D_s=|\Omega_c|^2-(\omega+d_{32})(\omega+d_{42})$.

\begin{figure}
\centering
\includegraphics[width=1\textwidth]{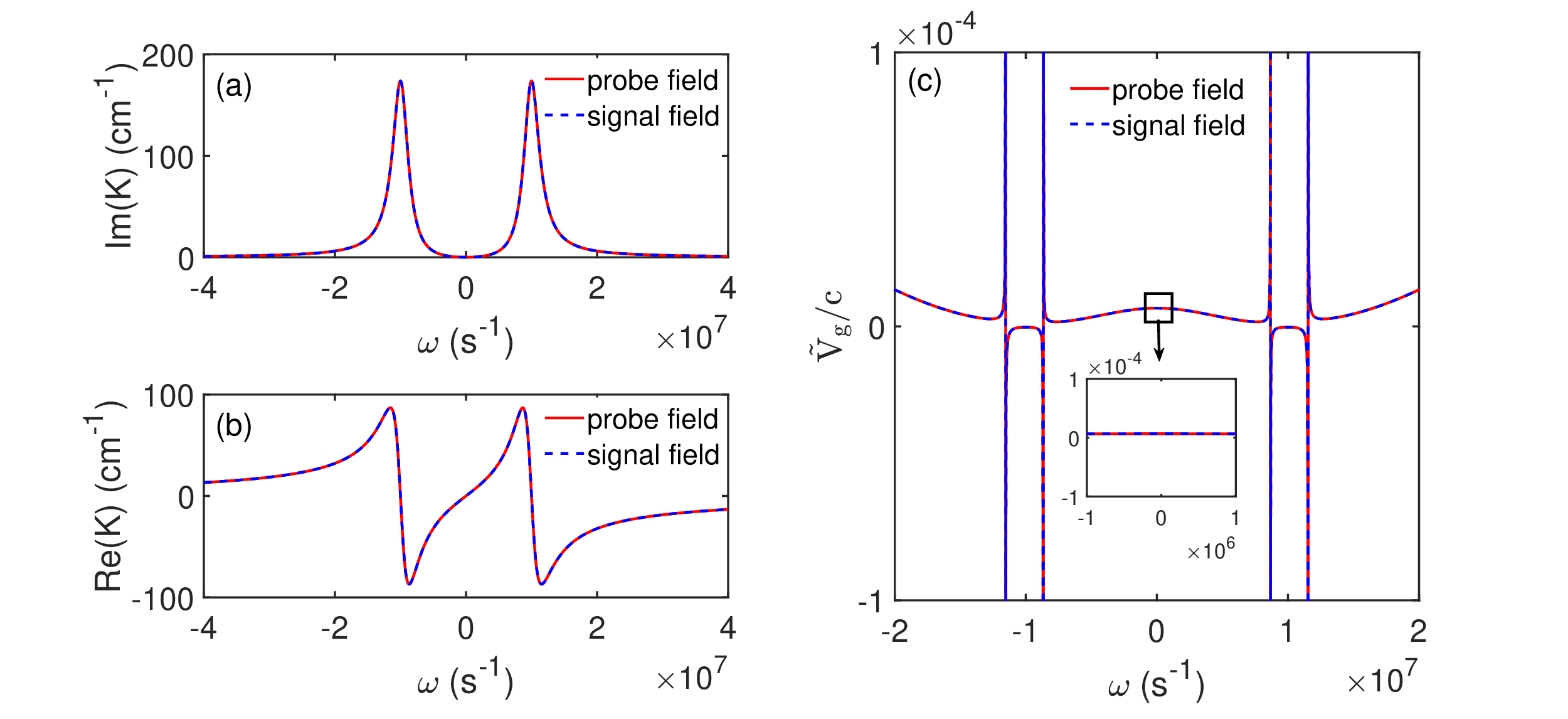}
\caption{The linear dispersion relations and group velocities of the probe field and signal fields. (a) The absorption spectrum ${\rm Im}(K_{p(s)})$ and (b) the dispersion relation ${\rm Re}(K_{p(s)})$ as functions of the frequency shift $\omega$. (c) The group velocity $\tilde{V}_g/c$ of the probe field and the signal field changes as functions of $\omega$. The inset panel shows the group velocity matching region. The red solid line and blue dotted line correspond to the probe field and the signal field, respectively.}~\label{Fig:2}
\end{figure}

$V_{g_1}={\rm Re}[\partial K_p/\partial\omega]^{-1}$ and $V_{g_2}={\rm Re}[\partial K_s/\partial\omega]^{-1}$ are the group velocities of probe field and signal field respectively. From Eqs.~(\ref{K_p}) and~(\ref{K_s}), we can find that the mainly difference of the linear dispersion relation $K_p$ and $K_s$ is induced by the parameters $\Delta_2$, $\kappa_{j4}$ and $\Gamma_{ij}$ ($i=3,4, j=1,2$), and the difference will cause the group velocities of the probe and signal field mismatch.
In order to study the XPM between the probe and signal fields, the group velocities of the two fields must be well matched. Thus,
in our following analyse, we choose these system parameters as $\Delta_2=0$, $\gamma_{13}=\gamma_{23}$ and $\gamma_{14}=\gamma_{24}$, which is a typical symmetrical double EIT scheme. Under this condition, the linear dispersion relations of probe field and signal field are almost the same, i.e., $K_p\approx K_s$, thus the group velocities are well matched as shown in Fig.\ref{Fig:2}.

Figure~\ref{Fig:2}~(a) and (b) show the imaginary part and real part of the linear dispersion relation $K_{p(s)}(\omega)$ as a function of the center frequency shift $\omega$. The system parameters are chosen from the D1-line transition of $^{87}$Rb atoms, with the energy levels selected as $|1\rangle=|5^{2}\textrm{S}_{1/2}, F=1, m_F=-1\rangle$, $|2\rangle=|5^{2}\textrm{S}_{1/2}, F=2, m_F=1\rangle$, $|3\rangle=|5^{2}\textrm{S}_{1/2}, F=2, m_F=0\rangle$, and $|4\rangle=|5^{2}\textrm{P}_{1/2}, F=2, m_F=0\rangle$~\cite{Chen 2019}. The decay rates are given by $\Gamma_4=5.75$ MHz, $\Omega_c=10$ MHz, $\Delta_2=\Delta_3=\Delta_4=0$, $\lambda_c=\lambda_p=\lambda_s=780$ nm. The parameters of waveguide are $\mu_1=0.83+0.0001i$, $\mu_2=1$, $\varepsilon_1=-31.14+0.32i$, $\varepsilon_2=1$, and $k(\omega_s)=k(\omega_p)=(8.17+0.0012i)\times10^4$ cm$^{-1}$. The electric dipole matrix elements are $|\textbf{p}_{14}|\simeq|\textbf{p}_{24}|=1.46\times10^{-27}$ C$\cdot$cm. We assume that atomic density $\mathcal{N}_a\approx1.10\times10^{11}$ cm$^{-3}$, then $\kappa_{14}\approx\kappa_{24}\approx1.0\times10^{9}$ cm$^{-1}$ s$^{-1}$.
As shown in Fig.~\ref{Fig:2}(a), we can see that the absorption doublet of the probe and signal fields are almost overlap under the double EIT effect, and as the dispersion properties shown in Fig.~\ref{Fig:2}(b), which means the double EIT effect can hugely suppress the absorption of the two fields, and can satisfy the group velocity matching condition at the same time.
In Fig.\ref{Fig:2}(c), the red solid line and blue dashed line represent the relationship between the group velocity $\tilde{V}_g/c$ ($\tilde{V}_g=Re(V_g)$) and the frequency shift $\omega$ of the probe light field and the signal light field, respectively. We can see the group velocity of the signal field match the group velocity of the probe field well, and with a subluminal ($10^{-4}c$).

\subsection{NONLINEAR PROPERTIES OF THE SYSTEM}
In this subsection, we study the nonlinear excitation of SPs in this system. In the third and fourth order solution of the MB equations, we can obtain the solvable conditions, where the probe field and signal field envelope function $F_1$ and $F_2$ satisfies
\begin{subequations}\label{W01}
\begin{eqnarray}
\label{W11}
&&i(\frac{\partial}{\partial x_2}+\frac{\partial}{\partial t_2}\frac{1}{V_{g_1}})F_1+\frac{c}{2\omega_p}\frac{\partial^2}{\partial y_1^2}F_1-W_{11}|F_1|^2F_1e^{-2\bar{\alpha_1}x_2}=0,\\
&&i(\frac{\partial}{\partial x_2}+\frac{\partial}{\partial t_2}\frac{1}{V_{g_2}})e^{i\theta_s}F_2+\frac{c}{2\omega_s}\frac{1}{n_{\rm eff}}\frac{\partial^2}{\partial y_1^2}e^{i\theta_s}F_2-W_{21}|F_1|^2e^{i\theta_s}F_2e^{-2\bar{\alpha_1}x_2}=0 \label{W21},
\end{eqnarray}
\end{subequations}
with $\bar{\alpha}=\epsilon^{-2}\alpha_1=\epsilon^{-2}\rm{Im}[K_p(\omega)]$, $W_{11}$ and $W_{21}$ being the nonlinear coefficients describing self phase modulation (SPM) of the probe field and XPM between the probe and signal field, respectively. The relation between the nonlinear coefficients and the self-kerr and cross-kerr susceptibilities are
\begin{subequations}\label{x}
\begin{align}
&\chi_{11}^{(3)}=\frac{2c}{\omega_p}\frac{|\mathbf p_{14}|^2}{\hbar^2}W_{11},\\
&\chi_{21}^{(3)}=\frac{2c}{\omega_s}\frac{|\mathbf p_{24}|^2}{\hbar^2}W_{21}.
\end{align}
\end{subequations}

In general, the coefficients in Eq.(\ref{W01}) are complex, thus, the system does not allow stable local nonlinear solutions. Fortunately, if the system works under the condition of double EIT, the imaginary part of these coefficients can be much smaller than the real part. In addition to the parameters mentioned above, we choose the other parameters as $\Omega_c=1.0\times10^6$ Hz, $U_0=2.24\times10^8$ Hz, $R_y=107$~nm, $\Delta_2=1.0\times10^4$ Hz, $\Delta_3=1.0\times10^5$ Hz and $\Delta_4=8.0\times10^7$ Hz. We can obtain that $W_{11}\approx(5.09+0.025i)\times10^{-13} \rm{cm^{-1} s^2}$, $W_{21}\approx(5.18+0.017i)\times10^{-13}~\mathrm{cm^{-1} s^2}$. Based on Eq.~(\ref{x}), the self-kerr and cross-kerr susceptibilities $\chi_{11}^{(3)}=(2.42+0.012i)\times10^{-2}~\mathrm{cm^2 V^{-2}}$, $\chi_{21}^{(3)}=(2.46+0.008i)\times10^{-2}~\mathrm{cm^2 V^{-2}}$, respectively, which corresponding to giant Kerr effects. The typical diffraction length $L_{\rm Diff}=\omega_{p(s)}R_y^2/c$ of the system is approximately $9.24\times10^{-6}$cm, and the typical nonlinearity length $L_{\rm Nonl}=1/[{U_0^2{\rm Re(W_{11})}}]$ is approximately equal to $L_{\rm Diff}$. Thus, the diffraction effect can balance the nonlinearity effect of the system. And then we have the linear absorption lengths $L_{\rm Ap(s)}=1/{\rm Im}(K_{p(s)}+k_{p(s)})$ of the system, that are approximately $L_{\rm Ap}=0.0836$~cm, $L_{\rm As}=0.0835$~cm, we can get the condition $L_{\rm Ap(s)}\gg L_{\rm Nonl}\approx L_{\rm Diff}$.
$R_y$ and $U_0=1/\sqrt{L_{\rm Diff}{\rm Re}(W_{11})}$ are radius in the transverse direction and typical half-Rabi frequency of the probe field, respectively.
Under the above parameters condition, it is possible to obtain two-component soliton solutions of Eq.(\ref{W01})~\cite{Chen 2019}.

\subsection{Coupled NLSEs}
For the convenience of research, we transform the Eq.(\ref{W11}) and Eq.(\ref{W21}) into dimensionless form, which are the coupled nonlinear Schr\"{o}dinger equations (NLSEs), and the expressions are as following
\begin{subequations}\label{W011}
\begin{eqnarray}
&&i(\frac{\partial}{\partial s}+\frac{1}{\lambda_1}\frac{\partial}{\partial \tau})u_1+\frac{1}{2}\frac{\partial^2}{\partial\xi^2}u_1-w_{11}u_1|u_1|^2=-iA_1u_1,\label{W111}\\
&&i(\frac{\partial}{\partial s}+\frac{1}{\lambda_2}\frac{\partial}{\partial \tau})u_2+\frac{1}{2}\frac{\partial^2}{\partial\xi^2}u_2-w_{21}u_2|u_1|^2=-iA_2u_2,\label{W211}
\end{eqnarray}
\end{subequations}
where the dimensionless physical quantity are defined as $u_j=\epsilon F_j/U_0e^{-\bar{\alpha}_j x_2}$, $s=x/L_{\rm Diff}$, $\tau=t/\tau_0$, $\lambda_j=V_{gj}\tau_0/L_{\textrm{Diff}}$, $\xi=y/R_y$, $w_{j1}=W_{j1}/{|W_{11}|^2}$ and $A_{1(2)}=L_{\rm Diff}\alpha_{1(2)}$. And $\tau_0$ is the pulse duration.
To solve the above equation, we assume $u_j(\tau,s,\xi)=g_j(\tau,s)v_j(\tau,\xi)$ with
\begin{equation}
g_j(\tau,s)= \frac{1}{\sqrt[4]{4\pi\rho_0^2}}e^{-(s-\lambda_j\tau)^2/{4\rho_0^2}},
\end{equation}
where $\rho_0$ is a free real parameter. After neglecting the small absorption coefficient $A_j$ and integrating the variable $s$, Eq.(\ref{W111}) and Eq.(\ref{W211}) are further simplified as
\begin{subequations}\label{W0111}
\begin{eqnarray}
&&\left(\frac{i}{\lambda_1}\frac{\partial}{\partial \tau}+\frac{1}{2}\frac{\partial^2 }{\partial \xi^2}\right) v_1-\frac{1}{2\sqrt{\pi}\rho_0}w_{11}|v_1|^2v_1=0,\label{W1111}\\
&&\left(\frac{i}{\lambda_2}\frac{\partial}{\partial \tau}+\frac{1}{2}\frac{\partial^2 }{\partial \xi^2}\right) v_2-\frac{1}{2\sqrt{\pi}\rho_0}w_{21}|v_1|^2v_2=0.\label{W2111}
\end{eqnarray}
\end{subequations}

And then, we will study the spatial focusing effect of SPs based on the above equation.

\section{SPATIAL FOCUSING OF SPS BASED ON XPM }{\label{Sec:4}}

Eq.~(\ref{W0111}) has many soliton pair solutions\cite{Kivshar 2003}, next, we choose three typical soliton pair solutions as the initial condition to study the XPM between the probe and signal fields, and spatial focusing of the signal field based on XPM.

(i) {\it Bright-bright solition pair.}~The expression of bright-bright solition pair read
\begin{equation}
%\begin{align}
v_j=\varsigma_j\mathrm{sech}[\varsigma_{jj}(\xi-\eta_j\tau-\xi_0)]e^{i[\eta_j\xi-(\eta_j^2-\varsigma_j^2)\tau/2-\varphi_0]},~(j=1,2)
%\end{align}
\end{equation}
\begin{figure}[htbp]
\includegraphics[width=0.32\textwidth]{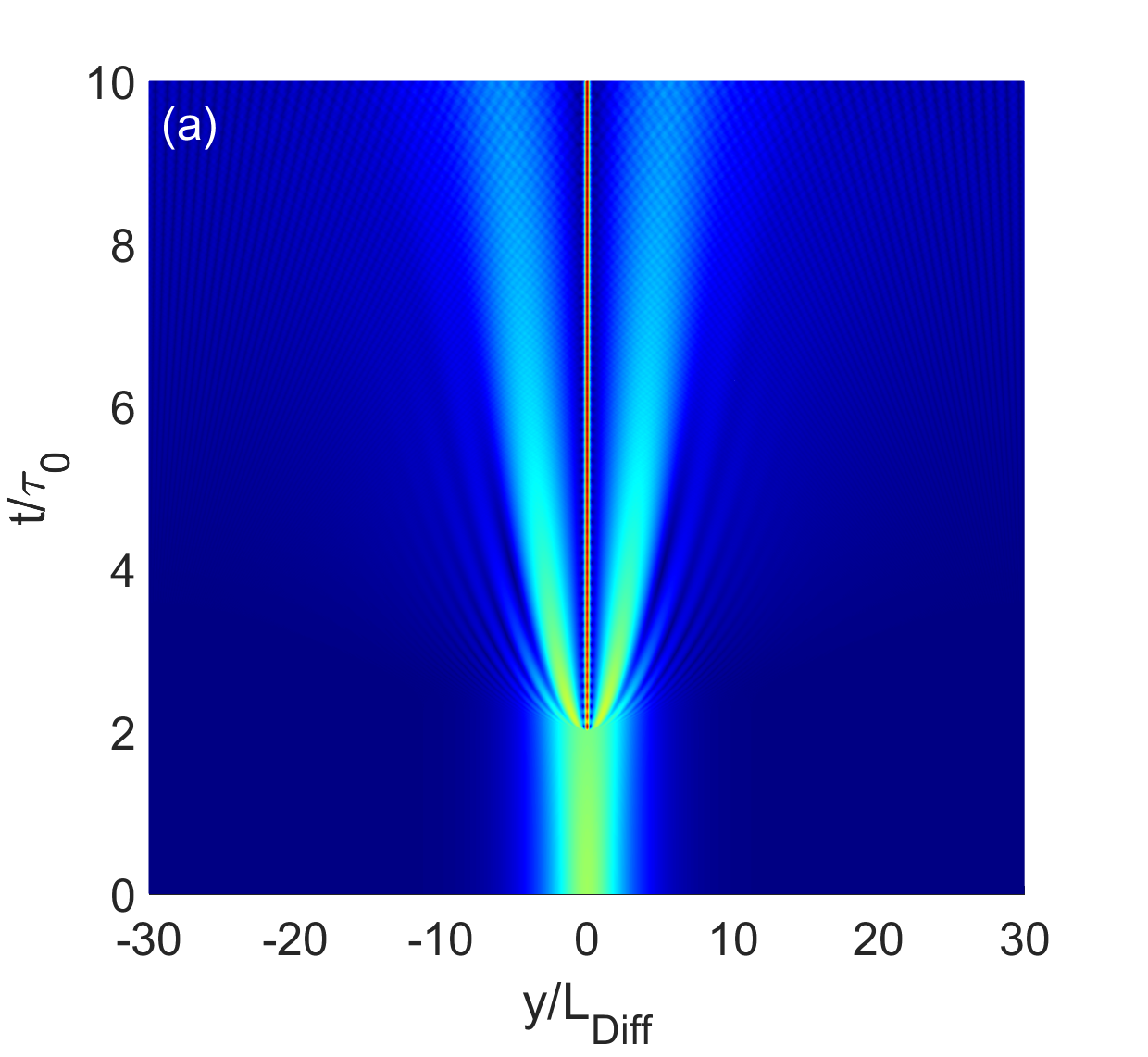}
\includegraphics[width=0.32\textwidth]{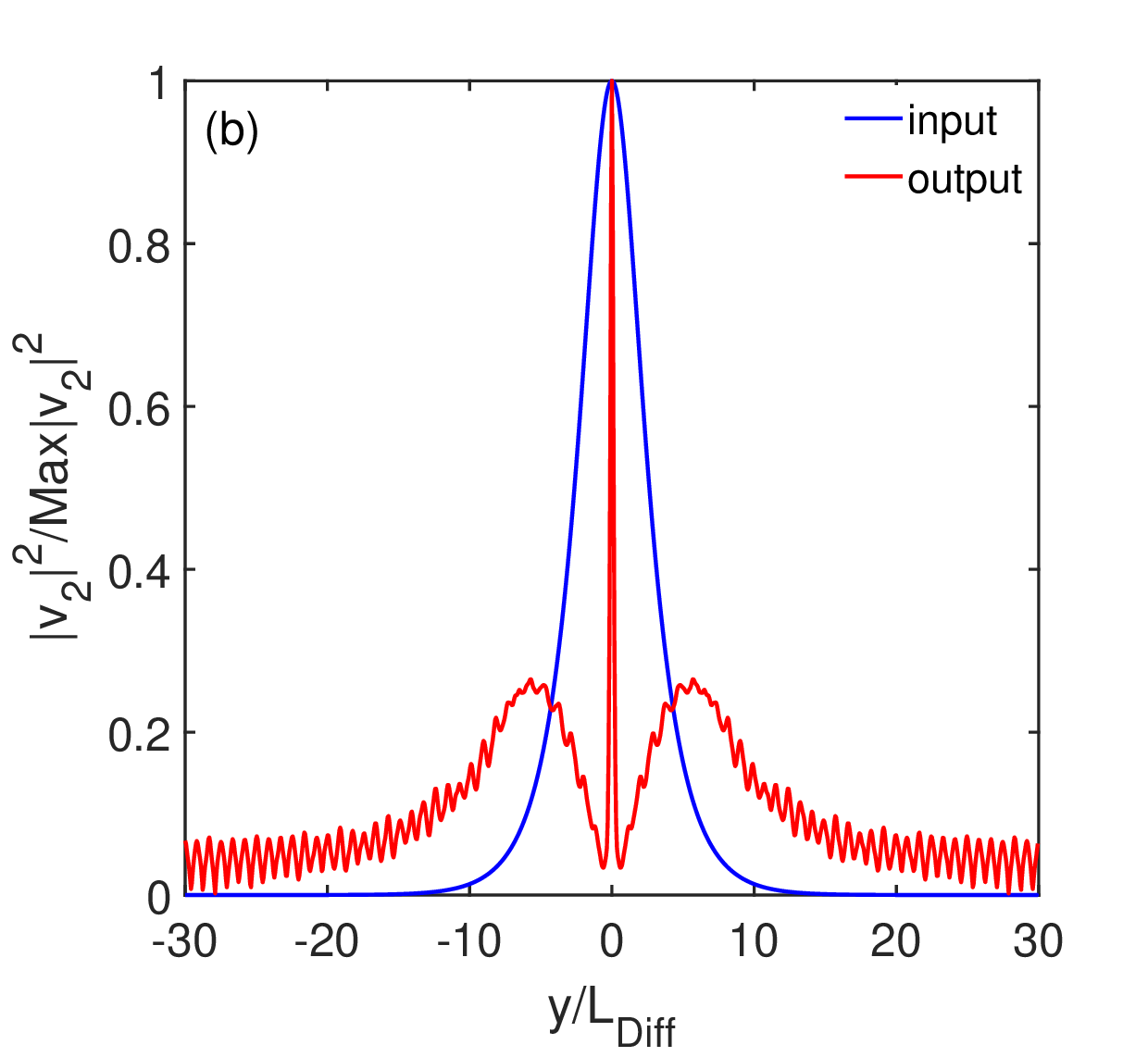}
\includegraphics[width=0.32\textwidth]{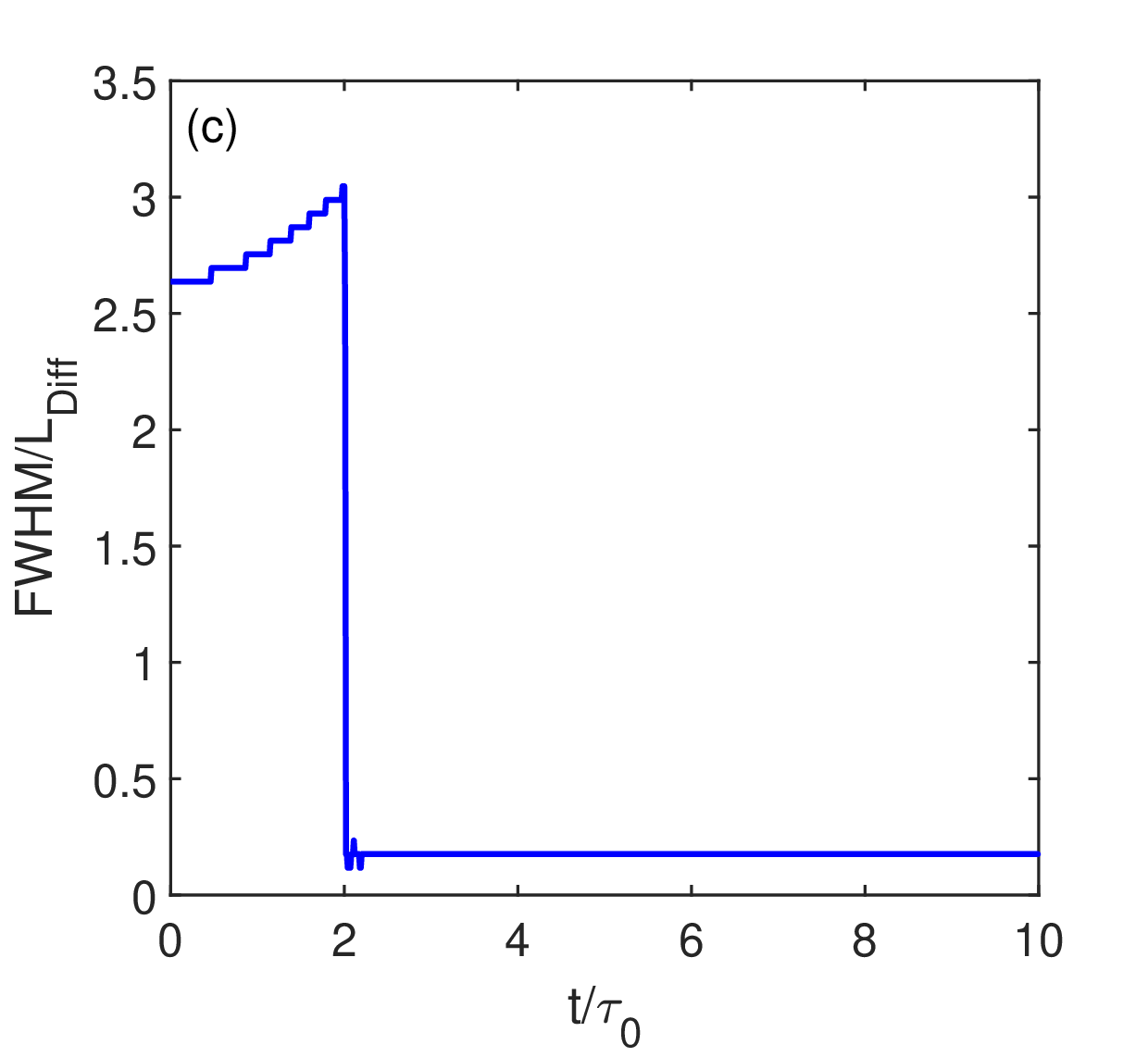}
\caption{(a) Time evolution of the bright SPs soliton $v_2$. After the probe soliton $v_1$ is turned on at the position of $t/\tau_0=2$, $v_2 $ will be obviously focused; (b) The input and output profile of the SPs soliton; (c) The FWHM of the SPs soliton changes with the propagation time $t/\tau_0$.}\label{Fig:3}
\end{figure}
where $\varsigma_j$, $\varsigma_{jj}$, $\eta_j$, $\xi_0$ and $\varphi_0$  are free real parameters~\cite{Chen 2019}.
For numerical simulation, we choose the following initial values of the free parameters as $\varsigma_1=8$, $\varsigma_2=1$, $\varsigma_{11}=6$, $\varsigma_{22}=0.6$, and the other are zeros.

In order to study the process of spatial focusing, we first input the SPs soliton $v_2$, and then, turn on the probe soliton $v_1$ at the position of $t/\tau_0=2$. The results are shown in Fig.\ref{Fig:3}. We can see that the SPs soliton first undergos a slight diffraction due to the absent of the term of XPM in Eq.~(\ref{W2111}), when the narrow probe soliton is turned on, the SPs soliton is focused to a very tiny area in a very shot response time, as shown in Fig.\ref{Fig:3}(a). The physical mechanism of such effect is that the narrow probe soliton modulates the profile of the SPs soliton in the overlap area via XPM. We can see that the full width at the half-maximum (FWHM) of the output profile of the SPs soliton is much small than that of the input one from Fig.\ref{Fig:3}(b). We also show the FWHM/$L_{\rm Diff}$ as function of the propagation time $t/\tau_0$ in Fig.\ref{Fig:3}(c). We can see that the FWHM of the initial SPs soliton is $2.2L_{\rm Diff}\sim203.28~$nm. After the spatial focusing, the curve dropped sharply, and the FWHM of the output SPs soliton is only $0.16L_{\rm Diff}\sim14.78~$nm, which is nearly $14$ times compressed. Thus, even the nanofocusing of SPs can be realized in our scheme.

(ii) {\it Bright-bright solition pairs.}
\begin{figure}[htbp]
\includegraphics[width=0.32\textwidth]{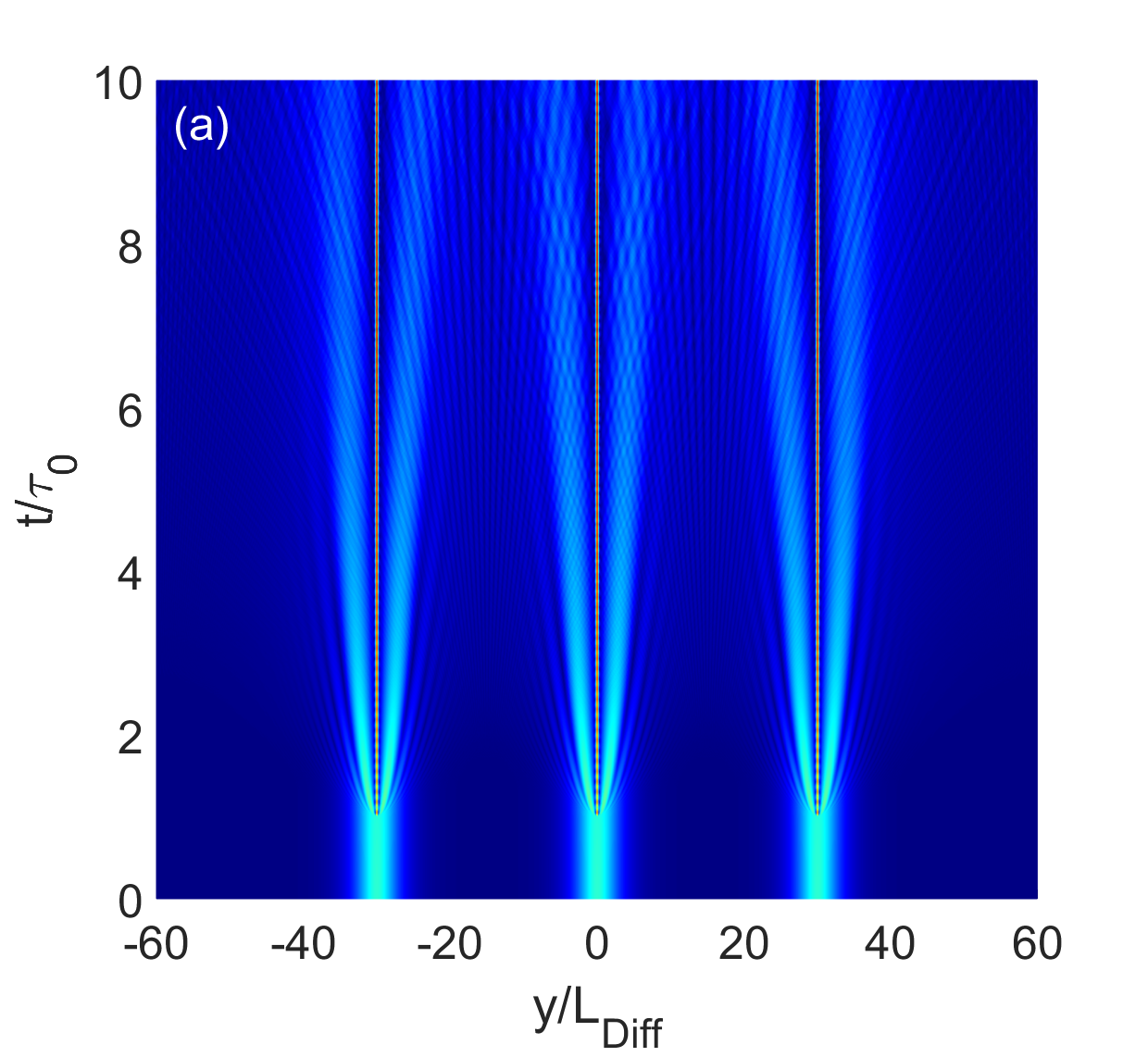}
\includegraphics[width=0.32\textwidth]{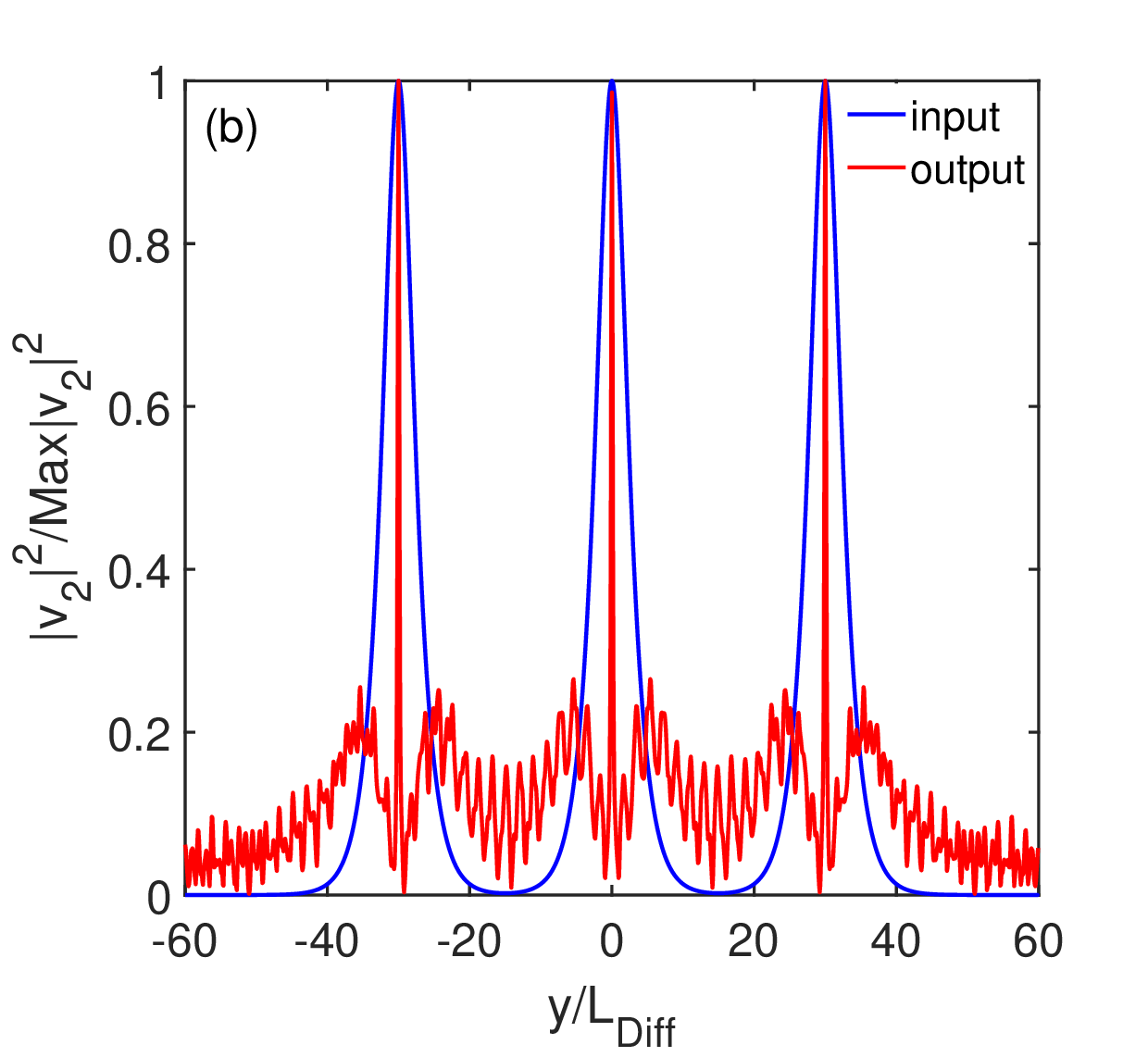}
\caption{(a) Time evolution of the multi SPs solitons $v_2$. After the probe soliton $v_1$ is turned on at the position of $t/\tau_0=1$, $v_2 $ will be obviously focused; (b) The input and output profile of the SPs solitons.}\label{Fig:4}
\end{figure}

Eq.~(\ref{W0111}) also has multi solition pairs solutions. Under the same condition as previous, we input three bright SPs solitons, and the initial positions are at $\xi_0=-30,~0~{\rm and}~30$, respectively. And then we turn on the probe solitons with the same form of solution at $t/\tau_0=1$, as shown in Fig.~\ref{Fig:4}(a). We find that such multi SPs solitons can also be focused with very narrow transverse width via XPM, the normalized input and output profiles are shown in Fig.~\ref{Fig:4}(b). The FWHM of the input field is about $3.3L_{\mathrm{Diff}}$. After the XPM, the pulse width of $v_2$ drops sharply and the output width is about $0.16L_{\mathrm{Diff}}$, which is compressed nearly $20$ times. Such results can be applied in the making of surface plasmon polaritons grating with high spectral resolution at micro/nano scale.

(iii) {\it Dark-dark solition pair.}
In order to verify the correctness of our theory, we also choose the dark solitons pair as the initial condition of our numerical simulation, the expressions of the dark solitons pair reads

\begin{equation}
%\begin{align}
v_j=\{\psi_j{i\mathrm{sin}{\phi_j+m_j\mathrm{cos}{\phi_j}\mathrm{tanh}[b_j(\xi-h\tau)]}\}e^{in_jc_j\xi+i[c_j^2/2+\chi_j]\tau}}, (j=1,2),
%&v_2=\psi_2{i\mathrm{sin}{\phi_2+\mathrm{cos}{\phi_2}\mathrm{tanh}[a(\xi-b\tau)]}\times e^{ic_2\xi+i[c_2^2/2+\chi]\tau}},
%\end{align}
\end{equation}
\begin{figure}[htbp]
\includegraphics[width=0.32\textwidth]{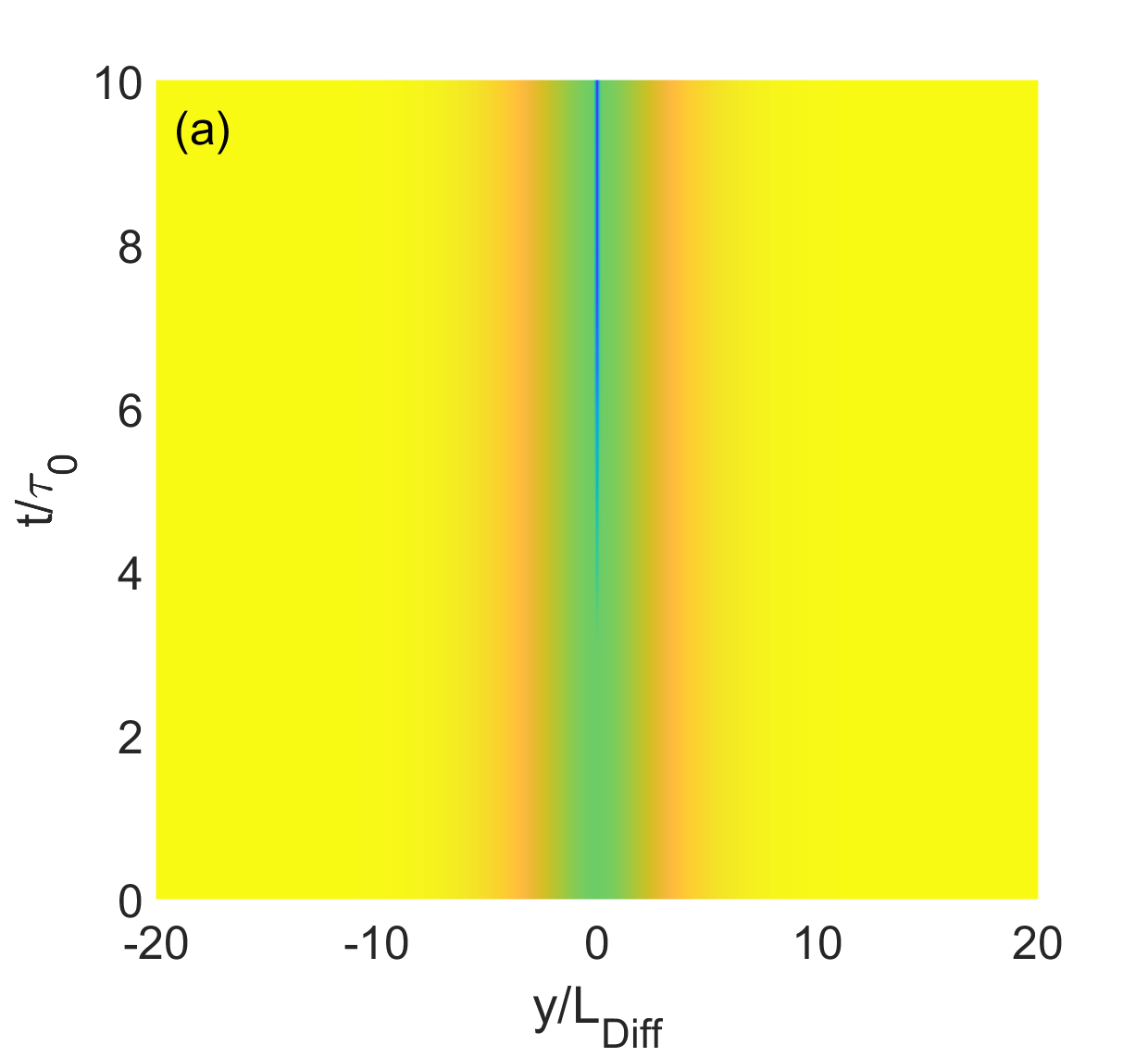}
\includegraphics[width=0.32\textwidth]{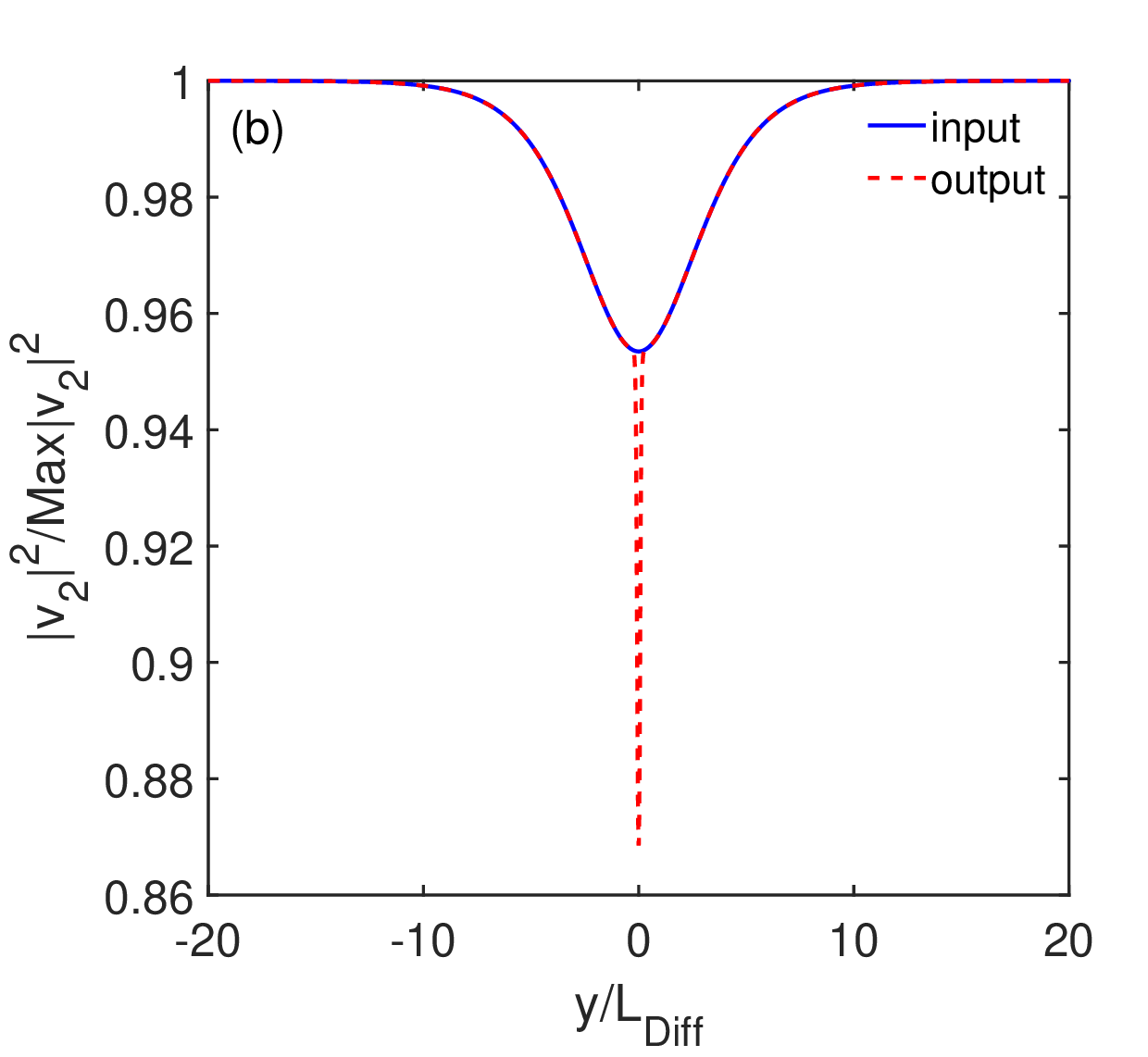}
\includegraphics[width=0.32\textwidth]{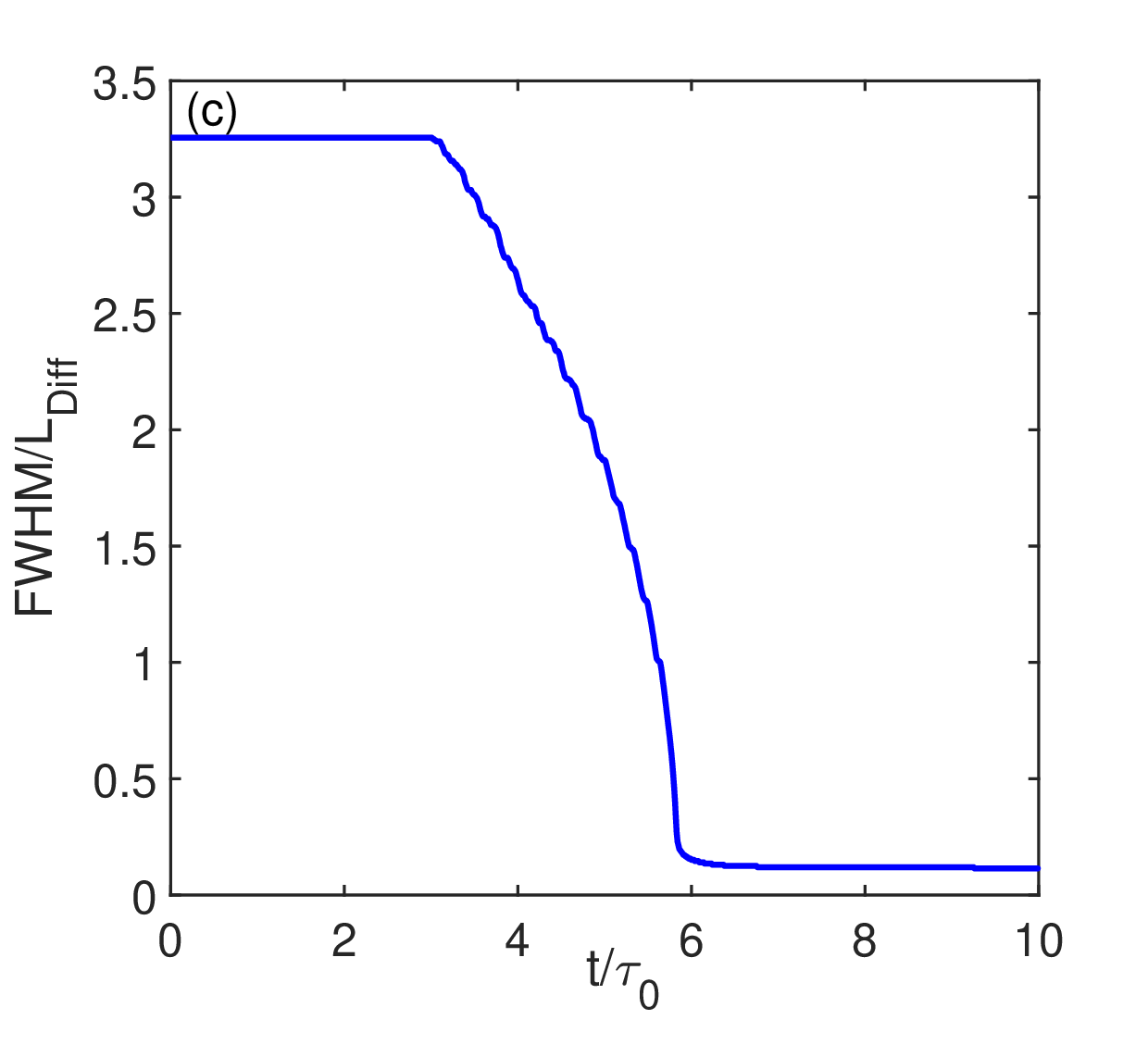}
\caption{(a) Time evolution of the dark SPs soliton $v_2$. After the probe soliton $v_1$ is turned on at the position of $t/\tau_0=3$, $v_2 $ will be obviously focused; (b) The input and output profile of the SPs soliton; (c) The FWHM of the SPs soliton changes with the propagation time $t/\tau_0$.}\label{Fig:5}
\end{figure}
where $\phi_j=\mathrm{arctan}{[(c_j-b_j)/a_j]}$, $\psi_j$, $a_j$, $b_j$, $\chi_j$ and $c_j$ are real parameters~\cite{Sheppard 1997}.
The initial values of the free parameters are $\psi_1=-8.14$, $a_1=3$, $b_1=18$, $c_1=2$, $m_1=6.51$, $n_1=0.4$, $\chi_1=-2$, $\psi_2=-1.20$, $a_2=\sqrt{0.8}$, $b_2=0.3$, $c_2=\sqrt{8}$, $m_2=1.07$, $n_2=0.28$, $\chi_2=-4$, and the other parameters are zeros.
The results are shown in Fig.\ref{Fig:5}, and with some differences comparing to results obtained in the bright one. We can see that the SPs soliton first propagates stably when the XPM is absent, and when turn on the dark probe soliton at $t/\tau=3$, a very narrow and deep dip appears in the overlap region between the signal and probe fields, {\it i.e.} spatial focusing, as shown in Fig.~\ref{Fig:5}(a). In Fig.~\ref{Fig:5}(b), we find that there are no higher harmonics beside the interacting region. The appearance of the narrow dip is due to the energy transfer in the XPM process. Shown in Fig.~\ref{Fig:5}(c) is the FWHM$/L_{\rm Diff}$ of the narrow dip as a function of $t/\tau_0$, we can see the FWHM of the SPs soliton drops slowly than that shown in Fig.~\ref{Fig:3}(c), which means the interacting time between the dark soliton is longer than that of the bright soliton.

\section{SUMMARY}{\label{Sec:5}}

In conclusion, a scheme based on XPM in double EIT is proposed to realize spatial focusing of SPs in a NIMM interface waveguide system. First, we obtain the low loss stable propagation of SPs, and the group velocities of the probe and signal fields are well matched under the double EIT condition in the linear region, and then, giant enhancement of the XPM can be obtained in the nonlinear region. Finally, the coupled NLSEs are derived in our system, by adopting the bright-bright soliton pair, multi bright solitons pairs and dark-dark soliton pair as the initial condition, we realize the spatial focusing of SPs solitons via XPM between the narrow optical soliton in free space and the SPs soliton, and even nanofocusing. We also find that the response time between the bright-bright soliton pair is much shorter than that of the dark one. These results not only provide a certain theoretical basis for realizing the active manipulation of SPs, but also have broad application prospects in the fields of micro-nano optics.

\begin{acknowledgments}
This work was supported by National Natural Science Foundation of China (NSFC) under Grant Nos. 11604185, 11704066, 11804196 and 11947072.
\end{acknowledgments}
%%%%%%%%%%%%%%%%%%%%%%%%%%%%%%%%%%%%%%%%%%%%%%%%

\section*{Appendix}
%%%%%%%%%%%%%%%%%%%%%%%%%%%%%%%%%%%%%%%%%%%%%%%%%%%%%%%%%%%%%%%%%%%%%%%%%%%%%%%%%
\appendix
\section{Expressions related to the electric field of SPs}
The permittivity ($\varepsilon_{1}$) and the permeability ($\mu_{1}$) of NIMM can be described by the Drude model, {i.e.}, $\varepsilon_{1}(\omega_s) = \varepsilon_{\infty}-\omega_e^2/\omega_s(\omega_s+i\gamma_e)$, $\mu_{1}(\omega_s) = \mu_{\infty}-\omega_m^2/\omega_s(\omega_s+i\gamma_m)$, where $\omega_{e,m}$ are electric and magnetic plasma frequencies of the NIMM, $\gamma_{e,m}$ are corresponding decay rates, and $\varepsilon_{\infty}$ and $\mu_{\infty}$ are background constants.

The decay coefficients along the $z$ direction read~
$k_j^2=k(\omega_s)^2-\varepsilon_{j}\mu_{j}\omega_s^2/c^2,$
where $j=1$ for the NIMM, $j=2$ for the atomic gas, and satisfies the relation $k_1\varepsilon_{2}=-k_2\varepsilon_{1}$, which gives the propagation constant of the SPs, i.e., $k(\omega_s)=\omega_s[\varepsilon_{1}\varepsilon_{2}(\varepsilon_{1}\mu_{2}-\varepsilon_{2}\mu_{1})/(\varepsilon_{1}^2-\varepsilon_{2}^2)]^{1/2}/c$.
The electric field of SPs in the atomic gas reads $\mathbf{E}_s(\mathbf{r},t)=\mathcal{E}_s(\mathbf{r},t)\mathbf{u}_s(z)\exp{[i(k(\omega_s) x-\omega_s t)]}+c.c.$, with the mode function being $\mathbf{u}_s(z)=-c[k(\omega_s)\mathbf{\hat{z}}-ik_2(\omega_s)\mathbf{\hat{x}})e^{k_2z}]/\varepsilon_{2}\omega_s$.
In our analysis, above system parameters are given by $\varepsilon_{\infty}=1$, $\mu_{\infty}=1$, $\omega_e=1.37\times10^{16}~{\rm s^{-1}}$, $\omega_m=2.45\times10^{15}~{\rm s^{-1}}$, $\gamma_e=2.73\times10^{13}~{\rm s^{-1}}$ (as for Ag), and $\gamma_m=\gamma_e/1000$. For a detailed derivation, one can refer to Ref.~\cite{Liu 2020}.
\section{Equations of motion for the density-matrix elements}\label{AppendixA}
The explicit Bloch equations describ the density matrix element $\sigma_{jl}$ $(j,l=1,2,3,4)$ under the interaction representation as follows:
\begin{subequations}\label{Bloch}
\begin{eqnarray}
&&i\frac{\partial}{\partial t}\sigma_{11}-i\Gamma_{14}\sigma_{44}+\Omega_p^*\sigma_{41}-\Omega_p\sigma_{41}^*=0,\label{A1} \\
&&i\frac{\partial}{\partial t}\sigma_{22}-i\Gamma_{24}\sigma_{44}+\zeta(z)^*e^{-i\theta_s^*}\Omega_s^*\sigma_{42}-\zeta(z)e^{-i\theta_s}\Omega_s\sigma_{42}^*=0 ,\label{A2} \\
&&i\frac{\partial}{\partial t}\sigma_{33}-i\Gamma_{34}\sigma_{44}+\Omega_c^*\sigma_{43}-\Omega_c\sigma_{43}^*=0,\label{A3} \\
&&i(\frac{\partial}{\partial t}+\Gamma_4)\sigma_{44}+\Omega_p\sigma_{41}^*+\zeta(z)e^{i\theta_s}\Omega_s\sigma_{42}^*+\Omega_c\sigma_{43}^*-\Omega_p^*\sigma_{41}-\zeta(z)^*e^{-i\theta_s^*}\sigma_{42}-\Omega_c^*\sigma_{43} =0,\label{A4}\\
&&(i\frac{\partial}{\partial t}+d_{21})\sigma_{21}+\zeta(z)^*e^{-i\theta_s^*}\Omega_s^*\sigma_{41}-\Omega_p\sigma_{42}^*=0,\label{A5} \\
&&(i\frac{\partial}{\partial t}+d_{31})\sigma_{31}+\Omega_c^*\sigma_{41}-\Omega_p\sigma_{43}^*=0,\label{A6} \\
&&(i\frac{\partial}{\partial t}+d_{32})\sigma_{32}+\Omega_c^*\sigma_{42}-\zeta(z)e^{i\theta_s}\Omega_s\sigma_{43}^*=0,\label{A7} \\
&&(i\frac{\partial}{\partial t}+d_{41})\sigma_{41}+\Omega_p(\sigma_{11}-\sigma_{44})+\zeta(z)e^{i\theta_s}\Omega_s\sigma_{21}+\Omega_c\sigma_{31}=0,\label{A8} \\
&&(i\frac{\partial}{\partial t}+d_{42})\sigma_{42}+\zeta(z)e^{i\theta_s}\Omega_s(\sigma_{22}-\sigma_{44})+\Omega_p\sigma_{21}^*+\Omega_c\sigma_{32}=0,\label{A9}\\
&&(i\frac{\partial}{\partial t}+d_{43})\sigma_{43}+\Omega_c(\sigma_{33}-\sigma_{44})+\Omega_p\sigma_{31}^*+\zeta(z)e^{i\theta_s}\Omega_s\sigma_{32}^*=0,\label{A10}
\end{eqnarray}
\end{subequations}
where $d_{jl}=\Delta_j-\Delta_l+i\gamma_{jl}$ $(j,l=1,2,3,4)$, the decoherence rate of change of the system is defined as $\gamma_{jl}=(\Gamma_j+\Gamma_l)/2$, where $\Gamma_j=\Sigma_{E_i<E_j}\Gamma_{ij}$.

\section{Solutions of the asymptotic expansion at the first and second orders}\label{AppendixB}
\subsection{Frist-order approximation}
\begin{subequations}
\begin{eqnarray}
&&\Omega_p^{(1)}=F_1e^{i\theta_1}=F_1e^{i[K_p(\omega)x_0-\omega t_0]},\\
&&\sigma_{31}^{(1)}=-\frac{\Omega_c^*\sigma_{11}^{(0)}}{D_p}F_1e^{i\theta_1},\\
&&\sigma_{41}^{(1)}=\frac{(\omega+d_{31})\sigma_{11}^{(0)}}{D_p}F_1e^{i\theta_1}.
\end{eqnarray}
\end{subequations}
The solution of other density matrix elements is zero, where $D_p=|\Omega_c|^2-(\omega+d_{31})(\omega+d_{41})$ and $\theta_1=K_p(\omega)x_0-\omega t_0$ are defined. $F_1$ is the pending envelope function of the probe field, which is related to the slow variables $y_1$, $t_2$ and $x_2$.

%According to the first-order approximate solution, we can obtain the linear dispersion relationship of the probe light field:
%\begin{equation}
%K_p(\omega)=\frac{\omega}{c}+\kappa_{14}\frac{(\omega+d_{31}^{(0)})\sigma_{11}^{(0)}}{|\Omega_c|^2-(\omega+d_{31}^{(0)})(\omega+d_{41}^{(0)})}.
%\end{equation}
\subsection{Second-order approximation}
\begin{subequations}
\begin{eqnarray}
&&\Omega_s^{(2)}=F_2e^{i\theta_2}=F_2e^{i[K_s(\omega)x_0-\omega t_0]},\\
&&\sigma_{11}^{(2)}=a_{11}^{(2)}|F_1|^2e^{-2\bar{\alpha_1}x_2},\\
&&\sigma_{22}^{(2)}=a_{22}^{(2)}|F_1|^2e^{-2\bar{\alpha_1}x_2},\\
&&\sigma_{32}^{(2)}=a_{32}^{(2)}\zeta_s(z)e^{i\theta_s}F_2e^{i\theta_2},\\
&&\sigma_{33}^{(2)}=a_{33}^{(2)}|F_1|^2e^{-2\bar{\alpha_1}x_2},\\
&&\sigma_{42}^{(2)}=a_{42}^{(2)}\zeta_s(z)e^{i\theta_s}F_2e^{i\theta_2}.
\end{eqnarray}
\end{subequations}
We define $D_s=|\Omega_c|^2-(\omega+d_{32})(\omega+d_{42})$ and $\theta_2=K_s(\omega)x_0-\omega t_0$. $F_2$ is the envelope function of the signal light field to be determined and related to the slow variables $y_1$, $t_2$ and $x_2$, and the coefficient $a_{jl}^{(2)}$ in the formula is:
\begin{subequations}
\begin{eqnarray}
&&a_{11}^{(2)}=-\frac{\sigma_{11}^{(0)}}{2D_p^*},\\
&&a_{22}^{(2)}=-\frac{\sigma_{11}^{(0)}}{2D_p^*},\\
&&a_{32}^{(2)}=-\frac{\Omega_c^*\sigma_{22}^{(0)}}{D_s},\\
&&a_{33}^{(2)}=-\frac{\sigma_{11}^{(0)}}{D_p^*},\\
&&a_{42}^{(2)}=-\frac{(\omega+d_{32})\sigma_{22}^{(0)}}{D_s}.
\end{eqnarray}
\end{subequations}
%According to the obtained second-order approximate solution, we can also derive the linear dispersion relationship of the weak signal light field, the expression is
%\begin{equation}
%K_s(\omega)=\frac{1}{n_{eff}}\frac{\omega}{c}+\kappa_{24}\frac{(\omega+d_{32}^{(0)})\sigma_{22}^{(0)}}{|\Omega_c|^2-(\omega+d_{32}^{(0)})(\omega+d_{42}^{(0)})}.
%\end{equation}
\subsection{Third-order approximation}
\begin{equation}
\sigma_{41}^{(3)}=i\frac{\partial}{\partial t_2}\frac{[|\Omega_c|^2+(\omega+d_{31})^2]\sigma_{11}^{(0)}}{D_p^2}\Omega_p^{(1)}+\frac{(\omega+d_{31})a_{11}^{(2)}}{D_p}|\Omega_p^{(1)}|^2\Omega_p^{(1)}.
\end{equation}
The explicit expression of the SPM cofficient $W_{11}$ in Eq.~(\ref{W11}) reads
\begin{equation}
W_{11}=-\kappa_{14}\frac{(\omega+d_{31})a_{11}^{(2)}}{D_p}=\kappa_{14}\frac{(\omega+d_{31})\sigma_{11}^{(0)}}{2|D_p|^2}.
\end{equation}
\subsection{Fourth-order approximation}
The explicit expression of the XPM cofficient $W_{21}$ in Eq.~(\ref{W21}) reads
\begin{equation}
W_{21}=-\kappa_{24}\frac{(\omega+d_{32})a_{22}^{(2)}}{D_s}=\kappa_{24}\frac{(\omega+d_{32})\sigma_{11}^{(0)}}{2D_sD^*_p}.
\end{equation}
%%%%%%%%%%%%%%%%%%%%%%%%%%%%%%%%%%%%%%%%%%%%%%%%%%%%%%%%%%%%%%%%%%%%%%%%%%%%%%%

\end{document}